# An Uncertainty-Guided Digital Twin Framework for Online Adaptive Proton Therapy in Head and Neck Cancer: A Feasibility Study

Yizhou Wu[1], Ryan J. Sanford[1], Huiqiao Xie[1], Jie Ding[1], Shupeng Chen[1], Tung-Ho Wu[2], Ping-Hsiu Wu[2], Justin Roper[1], Jun Zhou[1], Minglei Kang[3], Bill Stokes[1], Sibo Tian[1], David S. Yu[1], Xiaofeng Yang[4*], Chih-Wei Chang[1*]

[1]*Department of Radiation Oncology and Winship Cancer Institute, Emory University, Atlanta, GA 30308, USA*

[2]*Department of Radiation Oncology, Taipei Medical University Hospital, Taipei, 110301, Taiwan*

[3]*Department of Human Oncology, University of Wisconsin, Madison, WI 53792, USA*

[4]*Department of Radiation and Cellular Oncology, University of Chicago, Chicago, IL 60637, USA*

*Corresponding to: xfyang@uchicago.edu (XY) and chih-wei.chang@emory.edu (CC)

## Abstract

**Objective:** Head and neck (HN) proton therapy spans six to seven weeks of anatomical change, and offline replanning takes about a week. We present an uncertainty-guided digital twin (UGDT) framework that forecasts treatment-day anatomy before treatment and test whether it generates online adaptive proton therapy (APT) plans of clinical quality.

**Approach:** A library of 302 longitudinal deformations from 88 previously treated HN patients was transported onto each new patient's treatment planning computed tomography (TPCT) by two-step multi-atlas deformable image registration (DIR) built on a pretrained CT foundation model, giving about 284 predicted CTs (pdCTs) with contours per patient. The dispersion of the propagated clinical target volume (CTV) contours defined a patient-specific CTV robust margin for robust optimization. In ten patients, the quality assurance CT (QACT) that triggered a replan represented the treatment-day anatomy, and the physician-approved replan on it was the baseline. The pdCT most similar to the QACT (pdCT-H) and one from the lowest quartile (pdCT-L) were planned to within about 5% of the baseline plan quality score, forward-calculated on the QACT, and reoptimized on it to give the online APT plans.

**Main results:** pdCT plans scored within -0.7% (pdCT-H) and -1.0% (pdCT-L) of the baseline. Forward calculation on the QACT reduced the dose to 98% of the high-dose CTV (D98%) to 88.3% and 85.5%. After online reoptimization, D98% recovered to 98.3 ± 0.3% and 98.2 ± 0.3% against 98.5 ± 0.4% (baseline), spinal cord and brainstem doses stayed below tolerance, and plan quality scores were within -1.1% ($p = 0.19$) and -1.7% ($p = 0.01$) of the baseline for pdCT-H- and pdCT-L-initialized plans.

**Significance:** The UGDT framework produced online APT plans comparable in quality to physician-approved offline replans in HN proton therapy from anatomy forecast before treatment, a route from reactive offline replanning to anticipatory online adaptation.

**Key words:** digital twins, online adaptive proton therapy, head and neck cancer, deformable image registration, foundation model, contour uncertainty, plan quality

## 1. Introduction

Radiotherapy for HN cancer is delivered in 33 to 35 fractions over six to seven weeks, and the anatomy inside the immobilization mask does not stay fixed over that period. Serial imaging has shown that gross tumor volumes shrink by a median of about 70% by the end of a course, that parotid glands lose volume and migrate medially, and that these changes accumulate from week to week rather than fluctuating (Barker et al., 2004; Vásquez Osorio et al., 2008; Castadot et al., 2010). The proportion of oropharyngeal cancers attributable to human papillomavirus (HPV) has grown, and HPV-associated tumors are more radiosensitive than HPV-negative tumors because of impaired repair of DNA double-strand breaks (Ang et al., 2010; Kimple et al., 2013; Rieckmann et al., 2013); they regress faster and more completely during irradiation (Chen et al., 2013; Mohamed et al., 2018), so the anatomy of a later week departs further from the planning scan. Proton therapy is more sensitive to this departure than photon therapy, because the range of each pencil beam is set by the tissue upstream of the target, and a change in that tissue displaces the distal dose edge onto the parotid glands, oral cavity, brainstem or spinal cord (Lalonde et al., 2021; Huiskes et al., 2023). Proton centers therefore acquire QACTs during the course, recompute the plan of record on each one, and initiate an offline replan when a dose-volume criterion fails (Stanforth et al., 2022; Chang et al., 2024). In the institutional cohort of 88 HN proton patients analyzed in this work, 59 patients (67%) required at least one offline replan, and a median of 9 days (interquartile range 8 to 12) elapsed between the QACT that triggered a replan and the start of the replanned course. Each fraction delivered in that interval uses a plan optimized for an anatomy the patient no longer has, which is the motivation for adapting the plan online, with the patient on the treatment couch.

Online APT and its photon counterpart have been realized in several forms, and each of them begins after the treatment-day image is acquired. In photon therapy, cone-beam CT (CBCT)-guided online adaptive radiotherapy has been implemented for HN cancer with automated contouring and reoptimization, at session times near 20 min (Yoon et al., 2020; Blumenfeld et al., 2024; Shi et al., 2025), and daily adaptation has been demonstrated on a magnetic resonance linear accelerator (McDonald et al., 2021). In proton therapy, online workflows were described before they were deliverable (Albertini et al., 2020) and have since been realized as plan-parameter adaptation with Monte Carlo dose evaluation on the daily image (Botas et al., 2019), as automatic dose restoration (Borderías-Villarroel et al., 2022), and as a daily adaptive proton therapy service in which contouring, reoptimization, quality assurance and delivery fit into one treatment slot (Nenoff et al., 2019; Albertini et al., 2024; Choulilitsa et al., 2025). A precomputed plan library with plan-of-the-day selection has also been reported for proton therapy (Troost et al., 2022). Robust optimization absorbs part of the anatomical change inside the plan, but its protection extends only as far as the scenarios it was optimized against (Liu et al., 2012; Liu et al., 2013; Lalonde et al., 2021). The limitation shared by these approaches is their starting point. Because contouring and robust reoptimization of a HN plan with three dose levels and more than ten organs at risk (OARs) are the time-consuming steps, and because they can only begin once the anatomy of the day is known, the whole of the adaptation sits on the critical path of the treatment session. Models that anticipate anatomical change exist, as statistical deformation models (Söhn et al., 2005; Budiarto et al., 2011) and as generative or recurrent networks trained on HN cohorts (Yang et al., 2024; Burlacu et al., 2025), but they have been evaluated in the image domain, and none has been shown to yield a planning image on which a clinically acceptable proton plan can be built.

A digital twin (DT) is a virtual representation of a physical system that is updated by data from that system and that returns predictions to guide decisions about it (National Academies of Sciences and Medicine, 2024). In health care, DTs have been proposed for physiological monitoring, treatment selection and in silico trials (Katsoulakis et al., 2024; Sadée et al., 2025), and medical imaging is the data stream from which most DTs of individual anatomy are constructed (Zhao et al., 2025). For adaptive radiotherapy, the DT concept offers a potential paradigm shift: instead of reacting to the anatomy observed on the treatment day, the twin forecasts the anatomies a patient may present, is instantiated before the first fraction, and is synchronized with the patient by the daily image, so that the treatment-day decision reduces to selecting and refining a plan that already exists. Our group developed this concept for prostate stereotactic body radiotherapy, first as a CBCT-guided DT that derived patient-specific CTV setup uncertainty (Chang et al., 2025), then as a DT framework combining deep learning-based multi-atlas DIR with knowledge-based plan quality evaluation, which reached clinical-equivalent plan quality with a reduced reoptimization burden (Chang et al., 2026a; Chang et al., 2026b). For HN cancer, we recently showed that a prior patient's planning-to-QACT change, transported onto a new patient by a two-step foundation-model registration, predicts treatment-day anatomy more closely than the static planning CT (Wu et al., 2026a), and that the dispersion of the contours propagated across a library of prior patients quantifies the uncertainty of the forecast in six anatomical directions (Wu et al., 2026b). Both studies were confined to the image domain.

In this work we propose a UGDT framework for HN online APT and evaluate it in the dose domain. The framework transports anatomical changes observed in previously treated patients onto the new patient's TPCT by foundation model-based two-step multi-atlas DIR to generate pdCTs with propagated contours, and it derives from the contour uncertainty of the same ensemble a patient-specific CTV robust margin that enters robust optimization as the setup uncertainty. The question addressed is whether this framework can generate plans of clinical quality for HN cancer, a site with multiple prescription levels, many OARs and progressive anatomical change, and therefore a more demanding test than the prostate application in which the DT concept was first validated. The QACT acquired for clinical replanning serves as the reference image set, and the replan generated on it by a dosimetrist and approved by the treating physician serves as the clinical quality baseline. The initial pdCT-based plans are controlled to within about 5% of the plan quality score of that baseline; they are then evaluated on the QACT, which stands for the treatment-day anatomy, and reoptimized by the UGDT framework to give the online APT plans that are compared with the offline replan. Figure 1 gives an overview of the framework.

## 2. Materials and methods

### 2.1 Uncertainty-guided digital twin framework

The UGDT framework has an offline stage and an online stage, separated by the time at which they run and by the information available to them (Figure 1). The offline stage is completed before the first fraction. It transports a library of longitudinal deformations observed in previously treated HN patients onto the TPCT of the new patient, producing a patient-specific ensemble of pdCTs with propagated contours; it derives from the propagated CTV contours of that ensemble a direction-resolved CTV robust margin; and it generates treatment plans on selected pdCTs with the clinical objective template and the patient-specific margin. The online stage runs on the treatment day. The QACT is acquired in the treatment position and rigidly aligned on bone, each pdCT is scored against it by an image-only composite similarity, and the plan generated on the best-matching pdCT is evaluated on the QACT and reoptimized on it to produce the online APT plan, whose quality is graded by a knowledge-based plan quality score.

In the terms of the consensus definition of a DT (National Academies of Sciences and Medicine, 2024), the physical counterpart is the patient, observed on each treatment day through the QACT, and the virtual counterpart is the ensemble of pdCT image sets with their contours, margins and plans. Data pass from patient to twin through the QACT, which selects the ensemble member closest to the anatomy of the day, and from twin to patient through the reoptimized plan. The predictive element is the ensemble itself, which encodes anatomical change before it has occurred in this patient, and the uncertainty element is the spread of that ensemble, which is converted into the CTV robust margin. In this retrospective study the online stage was emulated with the QACT that had triggered a clinical offline replan, so that the online APT plan could be compared with the offline replan generated by a dosimetrist and approved by the treating physician on the same image. Two library members were carried into treatment planning for each patient: the pdCT of highest composite similarity to the QACT, denoted pdCT-H, and the pdCT at the 75th percentile of the same ranking counted from the top, that is, the highest-ranked member of the lowest quartile, denoted pdCT-L, which represents a poorly matched but non-degenerate library member. Planning both allows the dosimetric consequence of the image-similarity ranking to be examined: if the ranking is informative, a plan built on pdCT-H should hold up better on the QACT than a plan built on pdCT-L.

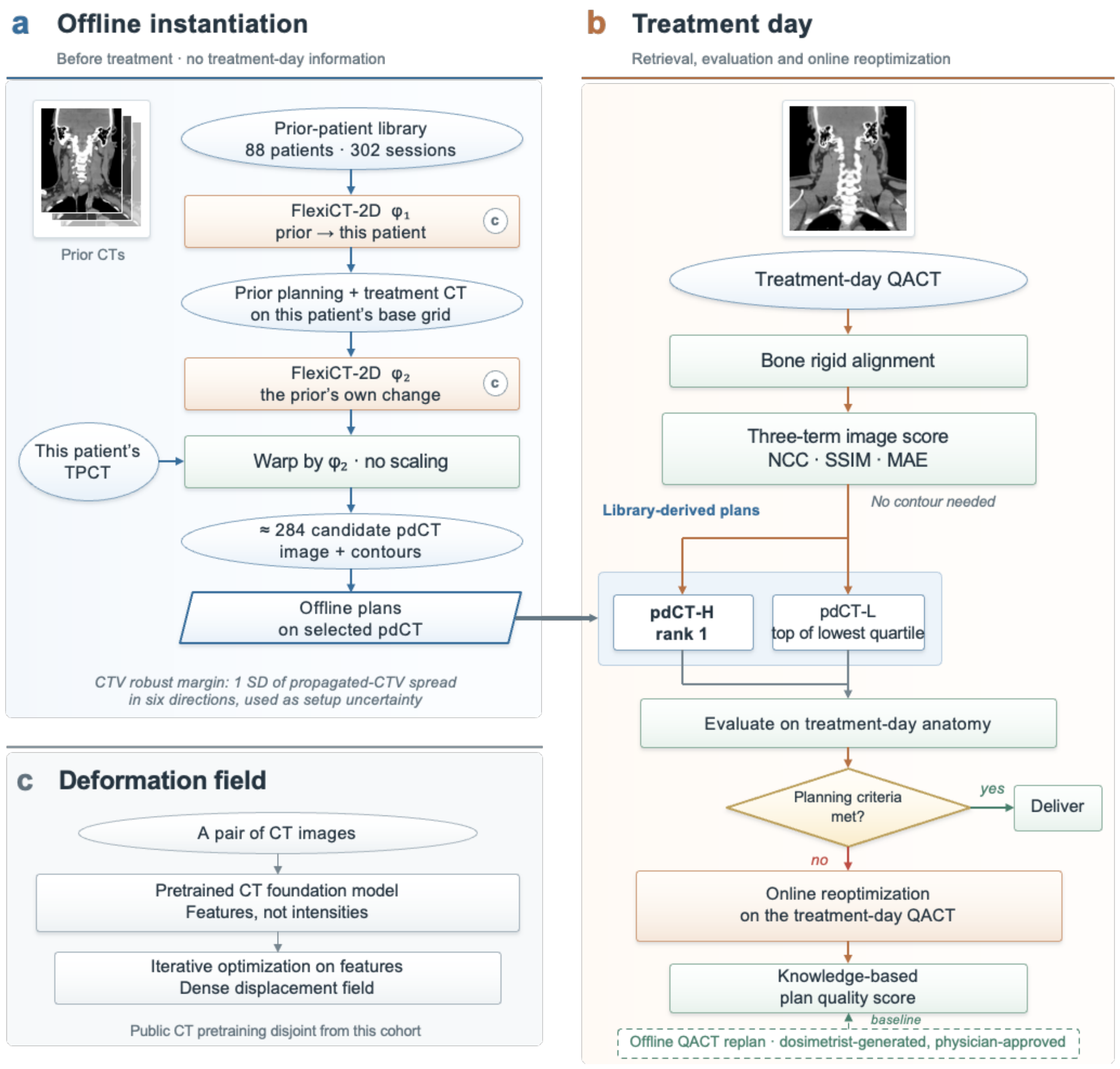

**Figure 1.** Overview of the UGDT framework for HN online APT. (a) Offline instantiation, completed before the first fraction. A library of 302 longitudinal deformations from 88 previously treated patients is transported onto the new patient's TPCT by two foundation-model registrations: φ1 carries a prior patient's planning and during-treatment CT onto the new patient's grid, and φ2, estimated on that transported pair, is the prior patient's own longitudinal change expressed in the new patient's frame. Warping the TPCT and its contours by φ2 gives one pdCT with propagated

contours per prior session, about 284 per patient. The spread of the propagated CTV contours across the ensemble, one standard deviation in each of six directions, is the patient-specific CTV robust margin used as the setup uncertainty in robust optimization, and offline plans are generated on the selected pdCTs. (b) Treatment day. The QACT is acquired in the treatment position and rigidly aligned on bone, each pdCT is scored against it by the three-term image-only composite of equation (6), and the plans generated on the highest-ranked pdCT (pdCT-H) and on the pdCT at the top of the lowest quartile (pdCT-L) are evaluated on the QACT and, when a planning criterion is violated, reoptimized on it to give the online adaptive plans. Their knowledge-based plan quality score is compared with that of the offline QACT replan generated by a dosimetrist and approved by the physician, which is the clinical baseline. (c) The registration engine: a pair of CT images is encoded by a pretrained CT foundation model and iterative optimization on those features returns a dense displacement field. Ellipses denote data, orange boxes registrations and optimization, green boxes other operations, the parallelogram a product carried forward to the treatment day, and the diamond the adaptation decision.

## 2.2 Patient data and clinical replanning context

The deformation library was drawn from 88 HN patients treated with intensity-modulated proton therapy (IMPT), each with a TPCT and three to seven verification QACTs acquired in the treatment position during the course; all sessions enter the library, giving 302 (prior patient, session) pairs, each contributing one observed longitudinal deformation. All images share a voxel spacing of $0.98 \times 0.98 \times 1.0$ mm³ on a $512 \times 512$ matrix, and all processing was performed at this resolution. The QACT is a diagnostic-quality CT acquired on a treatment day in the treatment position, the image against which proton range and delivered dose are verified at our institution (Stanforth et al., 2022; Chang et al., 2023). The same records describe the clinical context in which the QACTs were acquired (Figure 2). Treatment started a median of 16 days (interquartile range 14 to 19) after the TPCT. Offline replanning was carried out at least once in 59 of the 88 patients (67%) and twice or more in 27, giving 97 replans in the cohort, while 29 patients completed the course on the plan of record. For the 92 replans with recorded dates, the interval from the QACT that triggered the replan to the start of the replanned course was 9 days at the median (interquartile range 8 to 12). This interval is the elapsed cost of the offline path that the UGDT framework is designed to remove.

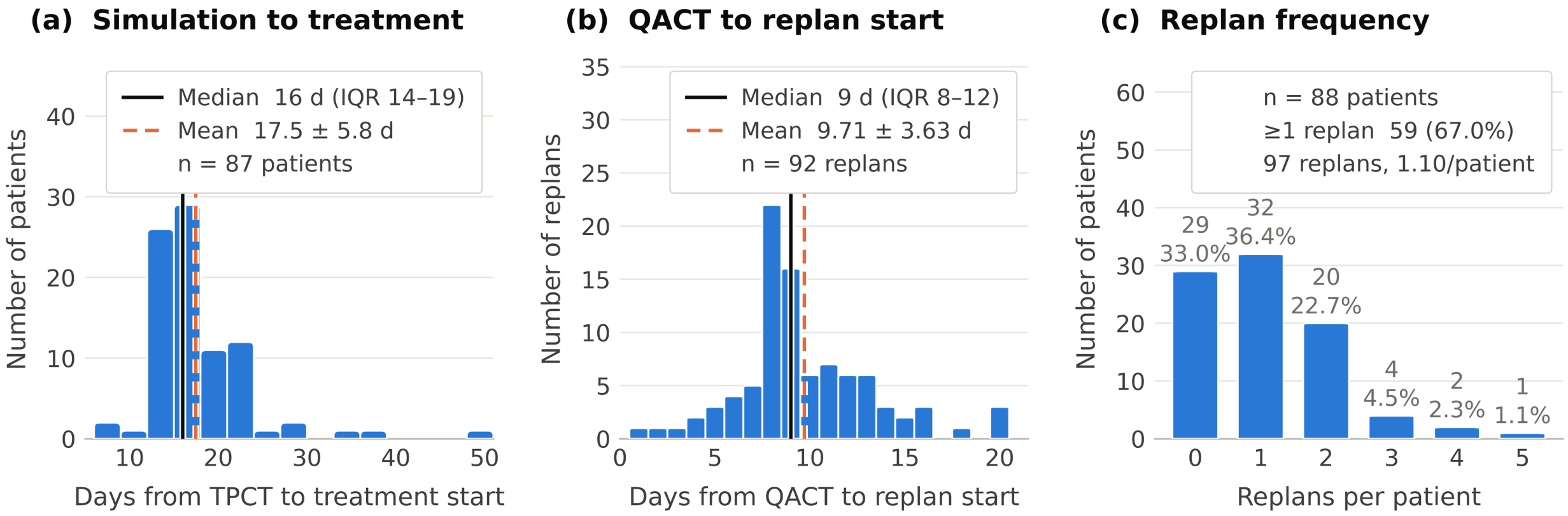

**Figure 2.** Treatment timing and offline replanning in the institutional HN proton cohort (n = 88). (a) Interval from the TPCT to the first fraction (n = 87 patients with recorded dates; median 16 d, interquartile range 14 to 19 d). (b) Interval from the QACT that triggered a replan to the start of the replanned course (n = 92 replans; median 9 d, interquartile range 8 to 12 d). (c) Number of offline replans per patient; 59 patients (67.0%) had at least one replan and 97 replans were generated in total. Solid line, median; dashed line, mean.

Ten of the 88 patients, identified as P01 to P10, form the evaluation cohort. Each had undergone a clinical offline replan, and the QACT that triggered that replan is the reference image set of this study: it stands for the treatment-day anatomy, and the replan generated on it by a dosimetrist and approved by the treating physician is the clinical quality baseline. For each evaluated patient the library was restricted to the other 87 patients, so that no patient's own longitudinal change contributed to its own prediction; after removal of the patient's own sessions and of transports that failed, 2,843 candidates remained across the ten patients, a mean of 284 per patient. Planning contours were delineated on the TPCT by the treating physician and propagated to each pdCT by the transported deformation field. On the QACT the CTVs and OARs were delineated by the treating physician. Patients received a simultaneous integrated boost with three prescription levels to a high-, intermediate- and low-dose CTV (CTV-H, CTV-M and CTV-L); an intermediate-dose volume was prescribed only in P01, P03 and P07. Doses are reported in Gy(RBE) with a constant relative biological effectiveness (RBE) of 1.1 and are written as Gy hereafter. CTV-H was prescribed 70 Gy in eight patients (P02 and P04 to P10) and 66 Gy in two (P01 and P03); CTV-L was prescribed 53.9 Gy in the 70 Gy patients and 54.12 and 54 Gy in P01 and P03; and CTV-M, where present, was prescribed 60.06, 60 and 60.2 Gy in P01, P03 and P07. All plans were delivered in 2 Gy fractions to CTV-H, that is 33 fractions for the 66 Gy prescriptions and 35 fractions for the 70 Gy prescriptions. The institutional planning criteria against which every plan was evaluated were, for each CTV, D98% at or above 98% of the prescription, and for the OARs a near-maximum dose (D0.03cc) below 45 Gy for the spinal cord and below 60 Gy for the brainstem, and a mean dose (Dmean) below 26 Gy for each parotid gland and below 35 Gy for the oral cavity. The parotid and oral cavity criteria are planning goals that cannot always be met when the structure overlaps a target. These criteria trigger adaptation in the online stage and define the scoring functions of the plan quality model.

### 2.3 Foundation model-based two-step multi-atlas deformable image registration

Let $I_B$ denote the TPCT of the patient being planned (the base patient) with planning contour set $S_B$. Prior patient *i* has a planning CT $I_i^P$ and verification CTs $I_i^{Q,k}$, where *k* indexes that patient's sessions; the term prior is used throughout for a previously treated patient whose observed change is transported. $R(F, M)$ denotes the registration model, which returns the deformation field $\varphi$ that maps moving image *M* onto fixed image *F*; $X \circ \varphi$ denotes image or contour set *X* resampled through $\varphi$, and a tilde marks an image that has been resampled onto the grid of the base patient. DIR is used here as correspondence machinery across patients and across time (Sotiras et al., 2013), and the transport below is defined for any registration model that returns a diffeomorphic field (Beg et al., 2005; Ashburner, 2007), so the registration engine can be exchanged or updated as better models or more data become available.

**Registration model.** Every registration was computed with FlexiCT-2D (Li et al., 2026), a vision-transformer CT foundation model pretrained by agglomerative self-supervised learning on 266,227 CT volumes from 56 public collections, applied through its deformable registration interface with the released weights. Each volume is encoded slice-wise; tokens from the last four transformer layers form a per-token descriptor that is projected onto a low-dimensional channel set by low-rank principal component analysis. Registration runs in two stages, a coupled-convex search at coarse resolution followed by dense refinement of the displacement field over 500 Adam iterations (learning rate 6.0) under a sum-of-squared-differences feature loss with a first-order diffusion penalty (weight 1.0); the remaining settings were identical for every pair. Optimizing on self-supervised features rather than raw intensities is what makes the cross-patient step tractable, since intensity-based registration between different patients is prone to local minima. The public pretraining data are disjoint from the institutional cohort.

**Two-step transport.** The transport takes two registrations and one warp. In the first step, one field per prior patient establishes cross-patient correspondence and carries both of that patient's images onto the grid of $I_B$:

$$\varphi_1^{(i)} = R\left(I_B, I_i^P\right), \qquad \tilde{I}_i^P = I_i^P \circ \varphi_1^{(i)}, \qquad \tilde{I}_i^{Q,k} = I_i^{Q,k} \circ \varphi_1^{(i)}. \tag{1}$$

Both images pass through the same field, so the transported pair differs only by the prior patient's own change between planning and session k; inter-patient anatomical difference has been removed. In the second step that change is estimated on the transported images, and is therefore expressed in the base patient's frame from the outset:

$$\varphi_2^{(i,k)} = R(\tilde{I}_i^{Q,k}, \tilde{I}_i^P). \tag{2}$$

The third step applies the transported change to the base patient's own planning image and contours,

$$I_{\text{pdCT}}^{(i,k)} = W\left(I_B, \varphi_2^{(i,k)}\right), \qquad S_{\text{pdCT}}^{(i,k)} = W\left(S_B, \varphi_2^{(i,k)}\right), \tag{3}$$

where $W(X, u)$ warps X by the displacement field u and contours are propagated by nearest-neighbor interpolation. The field is applied as estimated, with no per-patient scaling. The two registrations are complementary: $\varphi_1$ removes the prior patient's identity and $\varphi_2$ retains the prior patient's change, while applying $\varphi_2$ to $I_B$ keeps every candidate anchored to the anatomy of the patient being planned. Each (prior patient, session) pair gives one $\varphi_2$ and hence one pdCT with propagated contours; all transports were computed once per patient, offline, on a single graphics processing unit.

**Conjugation and vector reorientation.** Equation (2) is, to the accuracy of the registrations, a conjugation of the prior patient's own longitudinal deformation $\psi_i^{(k)}$ by the cross-patient map,

$$\varphi_2^{(i,k)} \approx \left(\varphi_1^{(i)}\right)^{-1} \circ \psi_i^{(k)} \circ \varphi_1^{(i)}, \tag{4}$$

where $\varphi_1(i)^{-1}$ is the inverse of the cross-patient map. This is the standard operation for expressing a deformation observed in one coordinate frame in another (Arsigny et al., 2006). Expanding (4) to first order in the displacements gives

$$u_2(x) = [D\varphi_1(x)]^{-1} u_\psi(\varphi_1(x)), \tag{5}$$

where $x$ is a point in the frame of the base patient, $u_2$ and $u_\psi$ are the displacement fields of $\varphi_2(i, k)$ and $\psi_i^{(k)}$, and $D\varphi_1$ is the Jacobian matrix of $\varphi_1(i)$. A transported displacement vector is therefore the prior patient's displacement, sampled at the corresponding point and reoriented by the inverse Jacobian of the cross-patient map. Pulling the vector back without the Jacobian factor drops the term $\nabla u_1 \cdot u_2$, where $u_1$ is the displacement field of $\varphi_1(i)$. The size of that term was measured on 80 transported fields, eight candidates for each of ten library patients, at a typical displacement magnitude of 6.22 mm: omitting the reorientation produced a median relative error of 36.3% (90th percentile 63.4%), a median angular deviation of 15.2° (90th percentile 30.8°) and an absolute error of about 2.26 mm. An error of a third of the displacement being transported places the reorientation at first order; it is part of the transport, not a refinement of it.

**Bone-based rigid pre-alignment.** Before any candidate is scored, each image is rigidly aligned to the QACT on bone, reproducing the setup correction that daily image guidance performs. The alignment uses a six-degree-of-freedom Euler transform (SimpleITK) with its rotation center at the QACT bone centroid. Bone images are formed from the Hounsfield unit (HU) image as clip(HU, 250, 1500) - 250; the metric is normalized cross-correlation with 25% sampling; optimization uses regular-step gradient descent (learning rate 2.0, minimum step $10^{-4}$, 400 iterations) over shrink factors 4, 2 and 1 with smoothing sigmas of 2.0, 1.0 and 0 mm; the metric region is the bounding box of the anatomy intersected with the QACT bone mask dilated by three voxels.

## 2.4 Patient-specific CTV robust margin from contour uncertainty

The transported ensemble carries more than a set of candidate images. Each pdCT arrives with a propagated CTV, and the ensemble of about 284 propagated CTVs for one patient is a sample of where that patient's target may lie on a treatment day, given the changes that previously treated patients underwent. The spread of that sample is a forecast uncertainty on the CTV boundary. It is uncertainty in the input to the forecast, namely which prior patient the current patient will follow, and not uncertainty in the parameters of a model; it is therefore patient-specific, and it is available before the first fraction. Its derivation and calibration are reported in a companion analysis (Wu et al., 2026b); here it is summarized to the extent needed to define the margin applied in treatment planning.

For each patient the propagated CTV-H of every candidate is represented by a signed distance field, and the ensemble mean contour is taken as the zero level set of the mean field. Each candidate is then described by its signed displacement from the mean contour along the outward normal of the mean contour, positive outside, area-weighted over the surface elements assigned to each of six direction bins. A surface element is assigned to the bin whose axis is closest to its outward normal, the axes being left, right, posterior, anterior, superior and inferior. This yields one signed displacement per candidate and direction, and the distribution over candidates in each direction is the contour uncertainty of that direction. The CTV robust margin of a patient in each direction is one standard deviation of the signed normal displacement over the ensemble in that direction, reported in millimeters with the results. Because the margin is derived from the same ensemble that produced the pdCTs, no additional imaging or training is required to obtain it.

In treatment planning the six margins were used as the patient-specific, direction-resolved setup uncertainty in robust optimization, in place of a population setup uncertainty. They are therefore an input to every plan evaluated in this study, and they replace a fixed isotropic margin by one that is smallest where the skull base holds the anatomy and largest where the neck contour and the low neck move over the course. The signed displacements themselves describe the ensemble and enter neither the candidate ranking nor any dose calculation.

## 2.5 Image similarity metrics and candidate ranking

After rigid alignment each candidate pdCT is compared with the QACT by three metrics that use the images alone, computed in three dimensions over the body. Normalized cross-correlation (NCC) measures voxel-wise intensity agreement up to a global scale and offset (Ayubi et al., 2024). The structural similarity index measure (SSIM) compares local means, variances and covariances and responds to structure at the scale of its window (Wang et al., 2004). Normalized mean absolute error (NMAE) measures intensity difference directly, computed on intensities normalized to [0, 1]. Each metric is mapped to [0, 1] with higher values indicating closer agreement, and the three are averaged with equal weight into the composite similarity $C_{img}$,

$$C_{img} = \tfrac{1}{3} \left[ (\mathrm{NCC} + 1)/2 + (\mathrm{SSIM} + 1)/2 + 1/(1 + \mathrm{NMAE}) \right] \quad (6)$$

The three terms cover voxel-wise intensity agreement, local structure and direct intensity error, and each is computable from the images alone. The weights are equal because no evidence exists that one of these metrics predicts dosimetric usefulness better than another, and because a weighting tuned on the evaluation cohort would enter the result it is meant to test. A learned perceptual term (Zhang et al., 2018) was not included because the available implementation is two-dimensional. Mutual information was not included because the pdCT and the QACT are both CT acquired under the same protocol, so the nonlinear intensity relationship that term tolerates is absent. The composite contains no contour-derived term. Ranking library members by image similarity to the target is the established rule for atlas selection in multi-atlas segmentation (Aljabar et al., 2009), and it serves the same purpose here: it predicts which library member will be most useful without measuring that usefulness directly. Because the composite is computed against the QACT, which is acquired before treatment delivery, the selection of pdCT-H is available at the time the adaptation decision is made and requires no delineation of the treatment-day image. Sorting the candidates of a patient from the highest to the lowest $C_{\text{img}}$, pdCT-H is the candidate at rank 1 and pdCT-L is the candidate at the 75th percentile counted from the top, that is, the highest-ranked member of the lowest quartile of the pool; it is worse than three quarters of the library but is not a failed transport.

### 2.6 Treatment planning

All plans were generated in RayStation 2023B (RaySearch Laboratories, Stockholm, Sweden) as IMPT plans with anterior-posterior (AP), right anterior oblique (RAO), left anterior oblique (LAO), right posterior oblique (RPO) and left posterior oblique (LPO) beams. Optimization was robust against setup and range uncertainties (Paganetti, 2012; Chang et al., 2020): the setup uncertainty was the patient-specific, direction-resolved CTV robust margin and the range uncertainty was 3.5%, and dose was computed with Monte Carlo simulation. A patient-specific template of optimization objectives was defined for the clinical offline replan of each patient and reused for every other plan of that patient, so that plans differ in the anatomy on which they were optimized and not in their objectives. Four plans enter the comparison for each patient. The offline QACT replan is the clinical plan generated by a dosimetrist on the QACT and approved by the treating physician, and it is the baseline of clinical plan quality. The pdCT-H and pdCT-L plans were optimized on the selected pdCT image sets with the propagated contours and the patient-specific margins; their plan quality score was controlled to within about 5% of the baseline score, so that the pdCT plans represent clinically acceptable plans rather than plans tuned to the anatomy of the day. Each pdCT plan was then forward-calculated on the QACT with the treatment-day contours to evaluate what a plan built on forecast anatomy delivers to the actual anatomy. Finally, each pdCT plan was reoptimized on the QACT with the same objective template, robustness settings and margins to give the online APT plan of the UGDT framework, initialized from pdCT-H or from pdCT-L. Dose evaluation before and after reoptimization used the QACT and the treatment-day contours.

### 2.7 Knowledge-based plan quality evaluation and statistical analysis

Plan quality is summarized by a single score built from the institutional planning criteria, following the ProKnow scoring approach (Nelms et al., 2012), which our group previously adopted for DT-based plan evaluation in prostate radiotherapy (Chang et al., 2026a). Each criterion defines one scoring function of its dose statistic: full marks are awarded when the statistic meets the criterion with margin, the score declines as the statistic approaches and crosses the tolerance, and it reaches zero at a value beyond which the plan would not be delivered. Target functions score D98% of each CTV level; OAR functions score the near-maximum dose of the spinal cord and brainstem and the mean dose of the parotid glands and oral cavity. The scoring functions are those of our previous study of online APT for HN cancer (Chang et al., 2024). The total plan quality score is the sum over functions, and a higher score denotes a plan closer to the intended quality. The relative score difference of a plan is (score - baseline score)/baseline score × 100%, with the offline QACT replan as the baseline. The score compresses many endpoints into one number and is used to compare plans; the individual dose-volume endpoints are reported alongside it. Statistical comparisons are paired within patient and use the two-sided Wilcoxon signed-rank test (Rosner et al., 2006), chosen for the paired design, the sample size of ten and the absence of a distributional assumption, at $\alpha = 0.05$ without multiplicity correction. The comparisons are the plan quality score of the pdCT-H and pdCT-L plans against the offline replan, the plan quality score of the online APT plans initialized from pdCT-H and from pdCT-L against the offline replan, and the scores of the two pdCT-initialized plan sets against each other.

## 3. Results

### 3.1 Ranking of predicted CTs for planning image set selection

For each of the ten patients the pdCT candidates instantiated offline were scored against the replanning QACT. Figure 3 shows the distribution of $C_{img}$ over each patient's pool, and Table 1 lists the composite and its three terms for pdCT-H and pdCT-L. Within a patient the distribution is narrow about its median, with interquartile ranges of about 0.01, and carries a lower tail extending 0.02 to 0.04 below the median, so that a small proportion of priors transport into an anatomy unlike the patient of the day. The level of the distribution differs between patients more than the spread within a patient: P03 and P06 have median composites near 0.955, whereas P07, P09 and P10 lie near 0.92 to 0.93, which reflects how far each patient's treatment-day anatomy departed from the library-derived candidates. Table 1 lists the three terms and the composite of the two candidates carried into planning. $C_{img}$ of pdCT-H ranged from 0.937 to 0.967 (mean 0.953) and that of pdCT-L from 0.915 to 0.945 (mean 0.930); the composite separating the two was 0.0217 at the median (range 0.0190 to 0.0289), and each of the three terms was lower for pdCT-L than for pdCT-H in every patient. This separation is the range of anatomical mismatch that the two planned arms span, and it is the quantity whose dosimetric consequence is examined below.

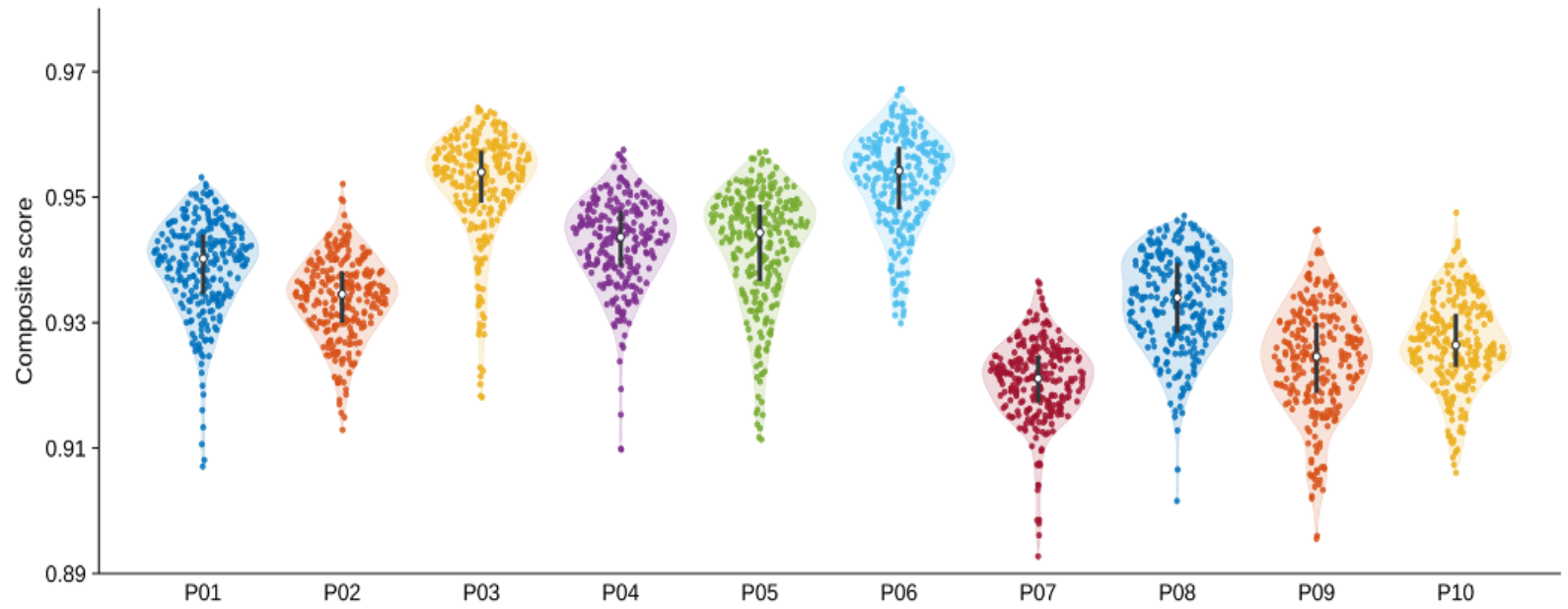

**Figure 3.** Composite image similarity $C_{img}$ between each patient's pool of pdCT candidates and the replanning QACT for the ten evaluated patients (P01 to P10). Each violin is the kernel density of $C_{img}$ of equation (6) over that patient's pool; every candidate is drawn as a dot with horizontal spread proportional to the local density, and the white marker and bar give the median and the interquartile range. The lowest tenth of each pool, which consists of failed transports, is not drawn.

**Table 1.** Image similarity between the replanning QACT and the two pdCTs carried into planning for the ten evaluated patients. pdCT-H is the pdCT of highest composite similarity to the QACT and pdCT-L the pdCT at the top of the lowest quartile of the same ranking. The three terms are given as they enter equation (6): NCC and SSIM directly, and NMAE as 1/(1 + NMAE), computed on min-max normalized images and therefore dimensionless. Every column is oriented so that a higher value is the closer match, and $C_{img}$ is their equally weighted mean.

| Patient | Planning image set | NCC | SSIM | 1/(1 + NMAE) | $C_{img}$ |
|---|---|---|---|---|---|
| P01 | pdCT-H | 0.944 | 0.813 | 0.981 | 0.953 |
| | pdCT-L | 0.901 | 0.747 | 0.973 | 0.932 |
| P02 | pdCT-H | 0.937 | 0.812 | 0.982 | 0.952 |
| | pdCT-L | 0.876 | 0.742 | 0.973 | 0.927 |
| P03 | pdCT-H | 0.962 | 0.853 | 0.986 | 0.964 |
| | pdCT-L | 0.926 | 0.788 | 0.979 | 0.945 |
| P04 | pdCT-H | 0.947 | 0.830 | 0.984 | 0.958 |
| | pdCT-L | 0.898 | 0.769 | 0.975 | 0.936 |
| P05 | pdCT-H | 0.947 | 0.829 | 0.984 | 0.957 |
| | pdCT-L | 0.896 | 0.751 | 0.975 | 0.933 |
| P06 | pdCT-H | 0.961 | 0.870 | 0.986 | 0.967 |
| | pdCT-L | 0.918 | 0.797 | 0.979 | 0.945 |
| P07 | pdCT-H | 0.908 | 0.763 | 0.974 | 0.937 |
| | pdCT-L | 0.862 | 0.699 | 0.966 | 0.915 |
| P08 | pdCT-H | 0.928 | 0.796 | 0.980 | 0.947 |
| | pdCT-L | 0.888 | 0.725 | 0.972 | 0.926 |
| P09 | pdCT-H | 0.957 | 0.815 | 0.949 | 0.945 |
| | pdCT-L | 0.916 | 0.707 | 0.936 | 0.916 |
| P10 | pdCT-H | 0.935 | 0.794 | 0.978 | 0.948 |
| | pdCT-L | 0.864 | 0.725 | 0.967 | 0.920 |

### 3.2 Patient-specific CTV robust margin

Figure 4 shows the ensemble of propagated CTV-H contours in the contour domain, as the signed normal displacement of each candidate from the ensemble mean contour in six directions, and Table 2 lists the CTV robust margin derived from it for each patient. The interquartile range of the displacement lies within about ±2.5 mm in every direction for every patient, the whiskers extend to between 3 and 5 mm, and the medians lie between 0 and 1.5 mm above the mean contour. Of the 17,058 candidate-direction points, 378 (2.2%) fall outside ±6 mm; these are the failed transports that also form the lower tail of Figure 3, and a candidate of that kind scores too poorly on the composite to be selected as pdCT-H and lies well below the rank that defines pdCT-L. The spread differs between patients more than between directions: P01, P04, P06 and P08 show the widest distributions and P03 and P05 the narrowest, in every direction. The resulting CTV robust margins range from 1.1 to 3.0 mm, with a mean of 1.8 mm over patients and directions. Averaged over patients, the margin is largest inferiorly (2.1 mm) and anteriorly (1.9 mm) and smallest superiorly (1.5 mm), so the margin is anisotropic and is widest where head and neck anatomy is least constrained by bone. Averaged over directions, the per-patient margin ranges from 1.3 mm in P05 to 2.4 mm in P10, and the ordering of patients by margin follows the ordering by ensemble spread in Figure 4. These margins entered robust optimization as the setup uncertainty of every pdCT plan and online APT plan reported below.

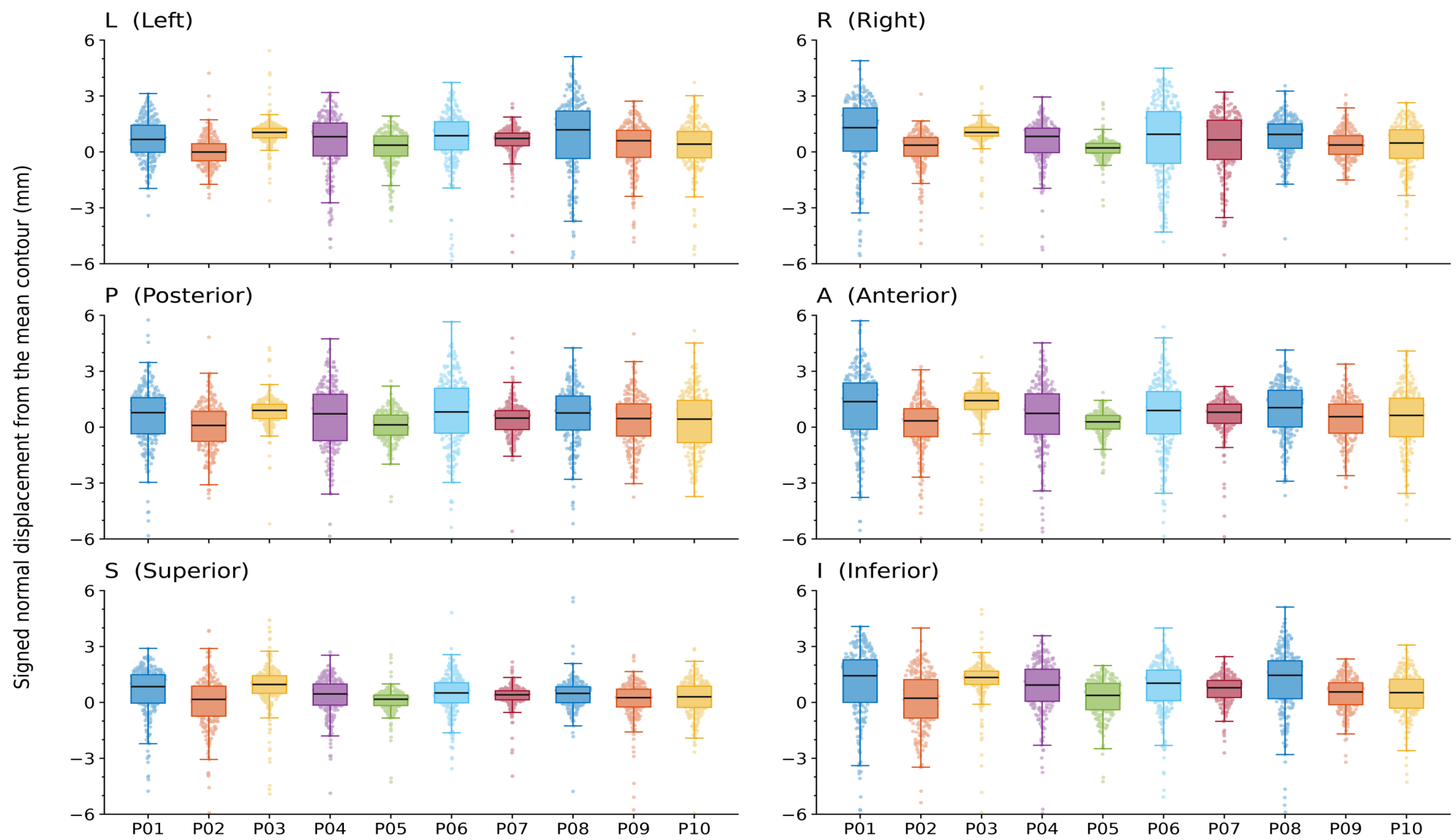

**Figure 4.** Range of anatomy spanned by the pdCT candidate pool on CTV-H, for the ten evaluated patients (P01 to P10). Each point is one candidate in one direction: the signed displacement of that candidate's propagated target volume from the ensemble mean contour, taken along the outward normal of the mean contour and area-weighted over the surface elements assigned to that direction, positive where the candidate lies outside the mean. The six panels are the left (L), right (R), posterior (P), anterior (A), superior (S) and inferior (I) direction bins. The box gives the interquartile range with whiskers at 1.5 times it, and every candidate is drawn as a dot behind it. The vertical axis is truncated to ±6 mm; 378 of 17,058 points (2.22%) fall outside it, almost all of them failed transports. One standard deviation of these distributions defines the patient-specific CTV robust margin of Table 2.

**Table 2.** Direction-resolved CTV robust margins applied in treatment planning, in millimeters, for each of the ten evaluated patients. The six directions are the left, right, posterior, anterior, superior and inferior axes of the patient frame. Each margin is one standard deviation of the signed normal displacement of the propagated CTV-H contours of the transported deformation ensemble and was used as the patient-specific setup uncertainty in robust optimization. The final row is the mean over the ten patients.

| Patient | Left | Right | Posterior | Anterior | Superior | Inferior |
|---|---|---|---|---|---|---|
| P01 | 1.7 | 3.0 | 1.9 | 2.2 | 1.5 | 2.8 |
| P02 | 1.5 | 1.5 | 1.6 | 1.6 | 1.4 | 1.8 |
| P03 | 1.4 | 1.4 | 1.4 | 1.5 | 1.2 | 1.5 |
| P04 | 2.0 | 1.9 | 2.1 | 2.2 | 1.8 | 2.3 |
| P05 | 1.3 | 1.2 | 1.4 | 1.5 | 1.1 | 1.5 |
| P06 | 1.5 | 1.7 | 1.6 | 1.8 | 1.4 | 1.8 |
| P07 | 1.7 | 2.3 | 1.7 | 1.8 | 1.3 | 1.8 |

| Patient | Left | Right | Posterior | Anterior | Superior | Inferior |
|---|---|---|---|---|---|---|
| P08 | 2.5 | 1.8 | 2.3 | 2.3 | 1.8 | 2.5 |
| P09 | 1.6 | 1.7 | 1.7 | 1.8 | 1.4 | 1.9 |
| P10 | 2.2 | 2.4 | 2.5 | 2.7 | 2.2 | 2.6 |
| **Mean** | **1.7** | **1.9** | **1.8** | **1.9** | **1.5** | **2.0** |

### 3.3 Initial treatment planning using pdCT

Table 3 summarizes the initial pdCT plans against the offline QACT replan, each evaluated on its own planning image. Target coverage was preserved on both predicted CTs. CTV-H D98% was 98.9 ± 0.5% for pdCT-H and 98.4 ± 0.7% for pdCT-L, compared with 98.5 ± 0.4% for the offline replan, and no pdCT plan fell below 97.4%. CTV-L D98% was at or above 97.0% in all plans and at or above 98.0% in every pdCT-H plan, and in the three patients with an intermediate-dose volume CTV-M D98% was at or above 98.7%. OAR doses stayed within the spinal cord and brainstem tolerances for every pdCT plan. The highest brainstem D0.03cc was 59.3 Gy in the pdCT-L plan of P07, whose offline replan was already 56.4 Gy, and the highest spinal cord D0.03cc was 38.8 Gy in the pdCT-L plan of P04. Averaged over the cohort, spinal cord D0.03cc was 5.8 Gy higher for pdCT-H and 7.3 Gy higher for pdCT-L than for the offline replan, brainstem D0.03cc was 5.0 and 6.7 Gy higher, and the parotid mean doses were 1.6 to 2.8 Gy higher; oral cavity Dmean differed by less than 2 Gy on average for both. These offsets reflect the residual anatomical difference between the forecast and the acquired anatomy on which the optimizer of each plan worked, and none changed the clinical acceptability of a plan. The plan quality score of the pdCT-H plans was within -0.7% of the baseline on average (range -5.1% to +1.7%); nine of ten were inside ±2%, and four scored higher than the offline replan. The pdCT-L plans scored -1.0% on average (range -5.3% to +3.1%), with six of ten inside ±2%. Neither difference from the baseline was significant ($p = 0.38$ for pdCT-H and $p = 0.25$ for pdCT-L), and the two pdCT plan sets did not differ from each other ($p = 0.57$). The largest reductions were in P06 (-5.1% and -5.2%), where both pdCT plans raised the parotid mean doses by 6 to 10 Gy relative to the offline replan. Plans generated on pdCT therefore met the plan quality standard of the offline QACT replans to within the intended 5% control, which establishes them as clinically acceptable starting points for the treatment-day evaluation that follows.

**Table 3.** Initial pdCT plans compared with the offline QACT replan, each evaluated on its own planning image, for the ten patients. QACT denotes the clinical offline replan optimized on the replanning QACT (baseline); pdCT-H and pdCT-L denote the plans generated on the pdCTs with high and low image similarity to the QACT. Target coverage is reported as D98% (% of prescription) for CTV-H, CTV-M and CTV-L; OAR dose as D0.03cc (Gy) for brainstem and spinal cord and Dmean (Gy) for the oral cavity and the left and right parotid glands. Score is the knowledge-based plan quality score; relative score difference is (pdCT - QACT)/QACT × 100%, positive values indicating a higher score than the offline replan. "–" indicates that the structure was not defined for that patient (CTV-M was prescribed only in P01, P03 and P07).

| Initial Planning | | CTV-H D98 (%) | CTV-M D98 (%) | CTV-L D98 (%) | Brainstem D0.03cc (Gy) | Spinal Cord D0.03cc (Gy) | Oral Cavity Dmean (Gy) | Left Parotid Dmean (Gy) | Right Parotid Dmean (Gy) | Score | Relative score difference (%) |
|---|---|---|---|---|---|---|---|---|---|---|---|
| P01 | QACT | 99.4 | 99.4 | 99.5 | 0.3 | 6.0 | 13.2 | 15.5 | 19.9 | 140.4 | |
| | pdCT-H | 98.9 | 99.3 | 99.1 | 0.6 | 6.1 | 12.6 | 18.1 | 20.0 | 139.3 | -0.8 |
| | pdCT-L | 99.1 | 99.1 | 99.2 | 0.5 | 6.6 | 12.9 | 19.8 | 19.8 | 139.5 | -0.7 |
| P02 | QACT | 98.1 | - | 98.1 | 8.1 | 14.3 | 20.8 | 17.4 | 38.5 | 119.2 | |
| | pdCT-H | 98.5 | - | 98.3 | 13.8 | 21.6 | 21.6 | 17.8 | 39.3 | 119.8 | 0.5 |
| | pdCT-L | 97.4 | - | 97.0 | 18.5 | 26.1 | 22.2 | 19.7 | 41.4 | 116.4 | -2.4 |
| P03 | QACT | 98.3 | 98.7 | 98.6 | 36.4 | 23.1 | 19.6 | 24.0 | 26.3 | 135.2 | |
| | pdCT-H | 98.8 | 98.9 | 98.7 | 42.5 | 34.9 | 18.8 | 26.5 | 25.8 | 133.9 | -1.0 |
| | pdCT-L | 97.5 | 98.7 | 98.4 | 44.6 | 33.2 | 18.9 | 23.6 | 25.0 | 135.2 | 0.0 |
| P04 | QACT | 98.0 | - | 97.7 | 9.7 | 19.7 | 19.7 | 22.2 | 13.1 | 125.4 | |
| | pdCT-H | 99.6 | - | 98.0 | 26.4 | 33.9 | 18.1 | 19.1 | 12.3 | 127.6 | 1.7 |
| | pdCT-L | 99.6 | - | 97.8 | 33.1 | 38.8 | 19.0 | 19.3 | 12.0 | 127.2 | 1.4 |
| P05 | QACT | 98.4 | - | 99.5 | 4.7 | 6.3 | 23.1 | 66.7 | 0.0 | 120.8 | |
| | pdCT-H | 98.9 | - | 99.2 | 6.6 | 6.5 | 30.2 | 67.7 | 0.1 | 121.1 | 0.2 |
| | pdCT-L | 98.0 | - | 99.5 | 5.4 | 6.3 | 32.7 | 68.6 | 0.0 | 120.6 | -0.2 |
| P06 | QACT | 98.9 | - | 98.6 | 3.7 | 13.9 | 9.7 | 12.3 | 20.2 | 127.5 | |
| | pdCT-H | 99.9 | - | 98.5 | 6.8 | 15.6 | 11.5 | 20.2 | 30.3 | 121.0 | -5.1 |
| | pdCT-L | 99.3 | - | 98.3 | 8.3 | 19.4 | 14.3 | 18.8 | 29.8 | 120.8 | -5.2 |
| P07 | QACT | 98.1 | 99.7 | 98.7 | 56.4 | 25.6 | 16.0 | 27.0 | 49.1 | 128.2 | |
| | pdCT-H | 98.4 | 99.9 | 99.0 | 58.7 | 31.3 | 20.4 | 27.3 | 54.0 | 128.7 | 0.3 |
| | pdCT-L | 98.4 | 99.9 | 99.0 | 59.3 | 29.8 | 17.3 | 23.2 | 53.1 | 132.2 | 3.1 |
| P08 | QACT | 98.2 | - | 98.5 | 7.0 | 8.8 | 9.6 | 48.5 | 16.3 | 119.8 | |
| | pdCT-H | 98.3 | - | 98.4 | 8.8 | 13.1 | 13.5 | 56.7 | 20.8 | 119.6 | -0.1 |
| | pdCT-L | 98.3 | - | 98.1 | 11.4 | 16.9 | 13.5 | 57.8 | 21.7 | 119.4 | -0.3 |
| P09 | QACT | 98.8 | - | 98.6 | 5.9 | 16.5 | 27.1 | 33.5 | 21.0 | 120.3 | |
| | pdCT-H | 99.0 | - | 98.5 | 8.5 | 21.9 | 27.7 | 32.9 | 26.2 | 118.2 | -1.7 |
| | pdCT-L | 98.0 | - | 98.2 | 11.1 | 23.6 | 27.7 | 33.5 | 28.6 | 114.0 | -5.3 |
| P10 | QACT | 98.3 | - | 99.5 | 7.8 | 8.3 | 11.8 | 24.1 | 26.6 | 125.1 | |
| | pdCT-H | 98.5 | - | 99.5 | 16.9 | 15.2 | 9.8 | 24.7 | 27.3 | 123.7 | -1.1 |
| | pdCT-L | 98.5 | - | 99.2 | 14.4 | 14.6 | 10.8 | 23.0 | 27.4 | 124.0 | -0.8 |

### 3.4 Online dose evaluation of pdCT plans on the QACT

Table 4 reports the forward calculation of each pdCT plan on the QACT, which represents the dose a plan built on forecast anatomy would deliver to the actual anatomy of the day if it were used without adaptation. Target coverage degraded markedly relative to the values on the predicted CT itself (Table 3). CTV-H D98% fell to a mean of 88.3% (median 91.0%, range 73.7% to 98.9%) for the pdCT-H plans and 85.5% (median 86.3%, range 73.4% to 97.3%) for the pdCT-L plans. Only one pdCT-H plan (P06) and three pdCT-L plans (P01, P03, P08) retained CTV-H D98% above 95%, and in six patients (P02, P03, P04, P05, P07, P09) at least one of the two plans dropped below 80%. CTV-L D98% was better preserved, with means of 93.7% (range 88.8% to 99.6%) for pdCT-H and 91.7% (range 86.3% to 97.1%) for pdCT-L, and CTV-M D98% in the three applicable patients ranged from 89.8% to 99.6%. On average the pdCT-H plans lost less coverage than the pdCT-L plans (10.6 against 13.0 percentage points of CTV-H D98%), but the advantage was not uniform: pdCT-H retained higher CTV-H D98% than pdCT-L in six patients and lower in four (P01, P03, P07, P08), with per-patient differences of up to 22 percentage points in either direction. OAR doses were less affected. Mean oral cavity dose was 20.4 and 19.9 Gy, and mean parotid doses 26.0 to 31.9 Gy, for pdCT-H and pdCT-L respectively, comparable to the values on the predicted CT. The brainstem D0.03cc remained below 60 Gy in every case (maximum 59.8 Gy, P07). Spinal cord D0.03cc rose to 41.3 Gy for the pdCT-H plan of P04 and exceeded the 45 Gy tolerance for the pdCT-L plans of P04 (46.8 Gy) and P07 (52.7 Gy). A plan optimized on a predicted CT, whatever its image similarity to the acquired anatomy, therefore does not retain clinical target coverage when applied directly to the QACT, and the online reoptimization step of the UGDT framework is required before delivery.

**Table 4.** Plans optimized on pdCT and forward-calculated on the replanning QACT without reoptimization. Dose-volume metrics for the plans generated on pdCT-H and pdCT-L, the pdCTs with high and low image similarity to the QACT, evaluated on the QACT with the treatment-day contours. Target coverage is reported as D98% (% of prescription) for CTV-H, CTV-M and CTV-L; OAR dose as D0.03cc (Gy) for brainstem and spinal cord and Dmean (Gy) for the oral cavity and the left and right parotid glands. "–" indicates that the structure was not defined for that patient (CTV-M was prescribed only in P01, P03 and P07).

| Online Evaluation | | CTV-H | CTV-M | CTV-L | Brainstem | Spinal Cord | Oral Cavity | Left Parotid | Right Parotid |
|---|---|---|---|---|---|---|---|---|---|
| | | D98 (%) | D98 (%) | D98 (%) | D0.03cc (Gy) | D0.03cc (Gy) | Dmean (Gy) | Dmean (Gy) | Dmean (Gy) |
| P01 | pdCT-H | 93.6 | 89.8 | 92.5 | 0.5 | 12.2 | 11.8 | 17.7 | 17.5 |
| | pdCT-L | 96.4 | 91.2 | 92.9 | 0.6 | 10.5 | 12.8 | 18.2 | 19.5 |
| P02 | pdCT-H | 83.2 | - | 88.8 | 17.2 | 30.7 | 18.8 | 18.1 | 44.2 |
| | pdCT-L | 73.4 | - | 91.0 | 21.3 | 31.3 | 19.1 | 19.3 | 45.0 |
| P03 | pdCT-H | 73.7 | 95.8 | 90.0 | 47.2 | 37.1 | 23.0 | 26.0 | 29.5 |
| | pdCT-L | 95.9 | 99.6 | 94.9 | 43.5 | 37.6 | 20.3 | 31.0 | 33.6 |
| P04 | pdCT-H | 91.8 | - | 92.7 | 26.5 | 41.3 | 22.6 | 18.5 | 12.3 |
| | pdCT-L | 79.0 | - | 86.3 | 37.1 | 46.8 | 20.5 | 19.4 | 12.4 |
| P05 | pdCT-H | 88.3 | - | 98.3 | 5.6 | 6.0 | 30.7 | 68.5 | 0.0 |
| | pdCT-L | 77.7 | - | 87.3 | 3.5 | 3.9 | 29.4 | 68.9 | 0.0 |
| P06 | pdCT-H | 98.9 | - | 97.7 | 6.6 | 18.6 | 11.7 | 19.9 | 28.4 |
| | pdCT-L | 87.3 | - | 87.8 | 5.8 | 14.6 | 10.6 | 18.4 | 30.6 |
| P07 | pdCT-H | 79.3 | 98.8 | 93.0 | 59.8 | 31.0 | 23.5 | 28.3 | 50.6 |
| | pdCT-L | 88.5 | 99.4 | 92.4 | 59.8 | 52.7 | 24.2 | 25.9 | 50.7 |
| P08 | pdCT-H | 90.2 | - | 92.1 | 12.6 | 25.1 | 16.3 | 58.0 | 22.0 |
| | pdCT-L | 97.3 | - | 97.1 | 21.5 | 36.7 | 16.8 | 58.0 | 22.4 |
| P09 | pdCT-H | 91.8 | - | 91.9 | 8.1 | 25.8 | 29.9 | 37.0 | 27.0 |
| | pdCT-L | 73.7 | - | 92.2 | 10.3 | 16.0 | 27.9 | 33.1 | 27.0 |
| P10 | pdCT-H | 92.3 | - | 99.6 | 22.1 | 15.5 | 15.8 | 25.7 | 28.2 |
| | pdCT-L | 85.3 | - | 95.3 | 31.5 | 30.2 | 17.1 | 26.6 | 28.2 |

### 3.5 Online adaptive treatment planning

Table 5 reports the online APT plans of the UGDT framework, in which each pdCT plan was reoptimized on the QACT, alongside the offline QACT replan. Reoptimization restored the target coverage that was lost when the pdCT plan was applied to the QACT without adaptation (Table 4). CTV-H D98% was 98.3 ± 0.3% for the plans initialized from pdCT-H and 98.2 ± 0.3% for those initialized from pdCT-L, compared with 98.5 ± 0.4% for the offline replan, and no online plan fell below 98.0%. CTV-L D98% was 98.6% and 98.4% on average for the two starting points (offline replan 98.7%), with a minimum of 97.6%, and CTV-M D98% in the three applicable patients was at or above 99.1% for every plan. OAR doses remained within tolerance. Brainstem D0.03cc exceeded 50 Gy only in P07, where the offline replan was already at 56.4 Gy and the online plans reached 57.1 and 57.8 Gy. Spinal cord D0.03cc was highest in P04 (34.3 and 43.8 Gy for the pdCT-H- and pdCT-L-initialized plans) and stayed below 45 Gy in all cases; averaged over the cohort it was 6.8 and 6.7 Gy higher than in the offline replan for the two starting points. Mean doses to the oral cavity and the parotid glands were 1.7 to 3.0 Gy higher than in the offline replan on average, with the largest increases in P02, P03 and P08.

The plan quality score of the online plans was within -1.1% of the offline replan on average when initialized from pdCT-H (range -6.6% to +1.0%) and within -1.7% when initialized from pdCT-L (range -6.5% to +0.4%). Eight of ten plans were inside ±2% for each starting point, and three pdCT-H-initialized plans (P02, P04, P10) scored higher than the offline replan. The difference from the offline replan was not significant for the pdCT-H-initialized plans ($p = 0.19$) and was significant for the pdCT-L-initialized plans ($p = 0.01$), although the two online plan sets did not differ significantly from each other ($p = 0.19$; pdCT-H-initialized higher in seven of ten patients). The two largest score reductions, in P03 (-6.6% and -6.5%) and in P10 for the pdCT-L-initialized plan (-4.1%), are accounted for by the higher brainstem, spinal cord and parotid doses in those rows. The online APT plans produced by the UGDT framework therefore matched the offline replan in target coverage and in the spinal cord and brainstem tolerances, with a small residual score deficit concentrated in a few patients, and the plans initialized from the high-similarity pdCT came closer to the baseline than those initialized from the low-similarity pdCT.

Figures 5 to 7 compare, for three representative patients (P01, P04 and P10), the offline replan computed on the QACT with the online APT plans initialized from pdCT-L and from pdCT-H, all computed on the same QACT. In each patient the dose colorwash of the pdCT-H-initialized plan (panel C) is visually close to the offline replan (panel A): the high-dose region conforms to CTV-H with the same shape and extent, and the intermediate isodose levels follow the same course through the neck. In the dose-volume histograms (DVHs, panel D) the target curves of all three plans are superimposed over their full range, consistent with the D98% values of Table 5. The differences between plans lie in the OARs. In P01 (Figure 5) the pdCT-L-initialized plan shifts the parotid curves toward higher dose, and both online plans place slightly more dose in the spinal cord than the offline replan (D0.03cc 7.3 and 8.3 against 6.0 Gy). In P04 (Figure 6) the spinal cord and brainstem curves of both online plans extend to higher dose than those of the offline replan, the pdCT-L-initialized plan more so (spinal cord D0.03cc 43.8 against 34.3 and 19.7 Gy; brainstem 29.5 against 22.9 and 9.7 Gy), whereas the parotid and oral cavity curves coincide. In P10 (Figure 7) the pdCT-L-initialized plan shifts the parotid and brainstem curves toward higher dose and departs from the offline replan in the oral cavity, while the pdCT-H-initialized plan reproduces the offline replan for every structure shown. The structures that separate the plans are those lying on steep dose gradients close to the targets, and in each of the three patients the plan initialized from the low-similarity pdCT departed further from the offline replan than the plan initialized from the high-similarity pdCT.

**Table 5.** Online adaptive plans generated by the UGDT framework, reoptimized on the replanning QACT from the pdCT plan, compared with the offline QACT replan, for the ten patients. QACT denotes the clinical offline replan optimized on the QACT (baseline). pdCT-H and pdCT-L denote the online adaptive plans initialized from the plan generated on the pdCT with high or low image similarity, respectively, and reoptimized on the QACT. Target coverage is reported as D98% (% of prescription) for CTV-H, CTV-M and CTV-L; OAR dose as D0.03cc (Gy) for brainstem and spinal cord and Dmean (Gy) for the oral cavity and the left and right parotid glands. Score is the knowledge-based plan quality score; relative score difference is (online plan - QACT)/QACT × 100%, positive values indicating a higher score than the offline replan. "–" indicates that the structure was not defined for that patient (CTV-M was prescribed only in P01, P03 and P07).

| Online Adaptation | | CTV-H D98 (%) | CTV-M D98 (%) | CTV-L D98 (%) | Brainstem D0.03cc (Gy) | Spinal Cord D0.03cc (Gy) | Oral Cavity Dmean (Gy) | Left Parotid Dmean (Gy) | Right Parotid Dmean (Gy) | Score | Relative score difference (%) |
|---|---|---|---|---|---|---|---|---|---|---|---|
| P01 | QACT | 99.4 | 99.4 | 99.5 | 0.3 | 6.0 | 13.2 | 15.5 | 19.9 | 140.4 | |
| | pdCT-H | 98.6 | 99.2 | 99.2 | 0.5 | 7.3 | 13.2 | 16.1 | 22.2 | 139.0 | -1.0 |
| | pdCT-L | 98.8 | 99.4 | 99.3 | 0.6 | 8.3 | 13.6 | 17.3 | 23.2 | 139.5 | -0.6 |
| P02 | QACT | 98.1 | – | 98.1 | 8.1 | 14.3 | 20.8 | 17.4 | 38.5 | 119.2 | |
| | pdCT-H | 98.5 | – | 98.1 | 12.5 | 33.7 | 24.0 | 18.7 | 46.3 | 119.6 | 0.3 |
| | pdCT-L | 98.1 | – | 98.4 | 20.1 | 31.6 | 24.7 | 20.7 | 48.0 | 119.5 | 0.2 |
| P03 | QACT | 98.3 | 98.7 | 98.6 | 36.4 | 23.1 | 19.6 | 24.0 | 26.3 | 135.2 | |
| | pdCT-H | 98.2 | 99.3 | 98.8 | 44.1 | 32.7 | 23.8 | 27.6 | 30.9 | 126.3 | -6.6 |
| | pdCT-L | 98.2 | 99.1 | 98.3 | 41.4 | 28.9 | 21.6 | 28.4 | 29.4 | 126.3 | -6.5 |
| P04 | QACT | 98.0 | – | 97.7 | 9.7 | 19.7 | 19.7 | 22.2 | 13.1 | 125.4 | |
| | pdCT-H | 98.7 | – | 98.0 | 22.9 | 34.3 | 19.6 | 21.5 | 13.3 | 126.7 | 1.0 |
| | pdCT-L | 98.1 | – | 97.9 | 29.5 | 43.8 | 19.9 | 21.7 | 13.2 | 125.9 | 0.4 |
| P05 | QACT | 98.4 | – | 99.5 | 4.7 | 6.3 | 23.1 | 66.7 | 0.0 | 120.8 | |
| | pdCT-H | 98.0 | – | 99.7 | 7.5 | 6.1 | 30.2 | 68.8 | 0.0 | 120.7 | -0.1 |
| | pdCT-L | 98.0 | – | 98.8 | 4.6 | 6.7 | 31.3 | 72.1 | 0.0 | 119.8 | -0.8 |
| P06 | QACT | 98.9 | – | 98.6 | 3.7 | 13.9 | 9.7 | 12.3 | 20.2 | 127.5 | |
| | pdCT-H | 99.0 | – | 98.9 | 5.2 | 17.3 | 12.3 | 16.9 | 25.4 | 126.5 | -0.8 |
| | pdCT-L | 98.4 | – | 98.0 | 5.0 | 17.3 | 12.1 | 16.6 | 24.8 | 125.5 | -1.5 |
| P07 | QACT | 98.1 | 99.7 | 98.7 | 56.4 | 25.6 | 16.0 | 27.0 | 49.1 | 128.2 | |
| | pdCT-H | 98.1 | 99.6 | 98.1 | 57.1 | 25.9 | 16.0 | 27.3 | 49.6 | 127.2 | -0.8 |
| | pdCT-L | 98.1 | 99.7 | 97.7 | 57.8 | 28.4 | 17.2 | 27.6 | 50.6 | 126.2 | -1.6 |
| P08 | QACT | 98.2 | – | 98.5 | 7.0 | 8.8 | 9.6 | 48.5 | 16.3 | 119.8 | |
| | pdCT-H | 98.1 | – | 98.5 | 7.6 | 20.7 | 13.3 | 53.8 | 19.7 | 119.6 | -0.1 |
| | pdCT-L | 98.0 | – | 98.1 | 6.1 | 14.7 | 12.4 | 53.4 | 18.6 | 119.2 | -0.5 |
| P09 | QACT | 98.8 | – | 98.6 | 5.9 | 16.5 | 27.1 | 33.5 | 21.0 | 120.3 | |
| | pdCT-H | 98.0 | – | 97.6 | 6.6 | 24.5 | 30.2 | 33.4 | 25.8 | 116.5 | -3.2 |
| | pdCT-L | 98.0 | – | 98.2 | 5.8 | 20.1 | 30.6 | 34.1 | 24.8 | 118.4 | -1.6 |
| P10 | QACT | 98.3 | – | 99.5 | 7.8 | 8.3 | 11.8 | 24.1 | 26.6 | 125.1 | |
| | pdCT-H | 98.1 | – | 99.4 | 9.2 | 7.5 | 10.6 | 23.6 | 25.9 | 125.6 | 0.4 |
| | pdCT-L | 98.0 | – | 99.1 | 12.0 | 9.3 | 11.6 | 25.9 | 28.6 | 119.9 | -4.1 |

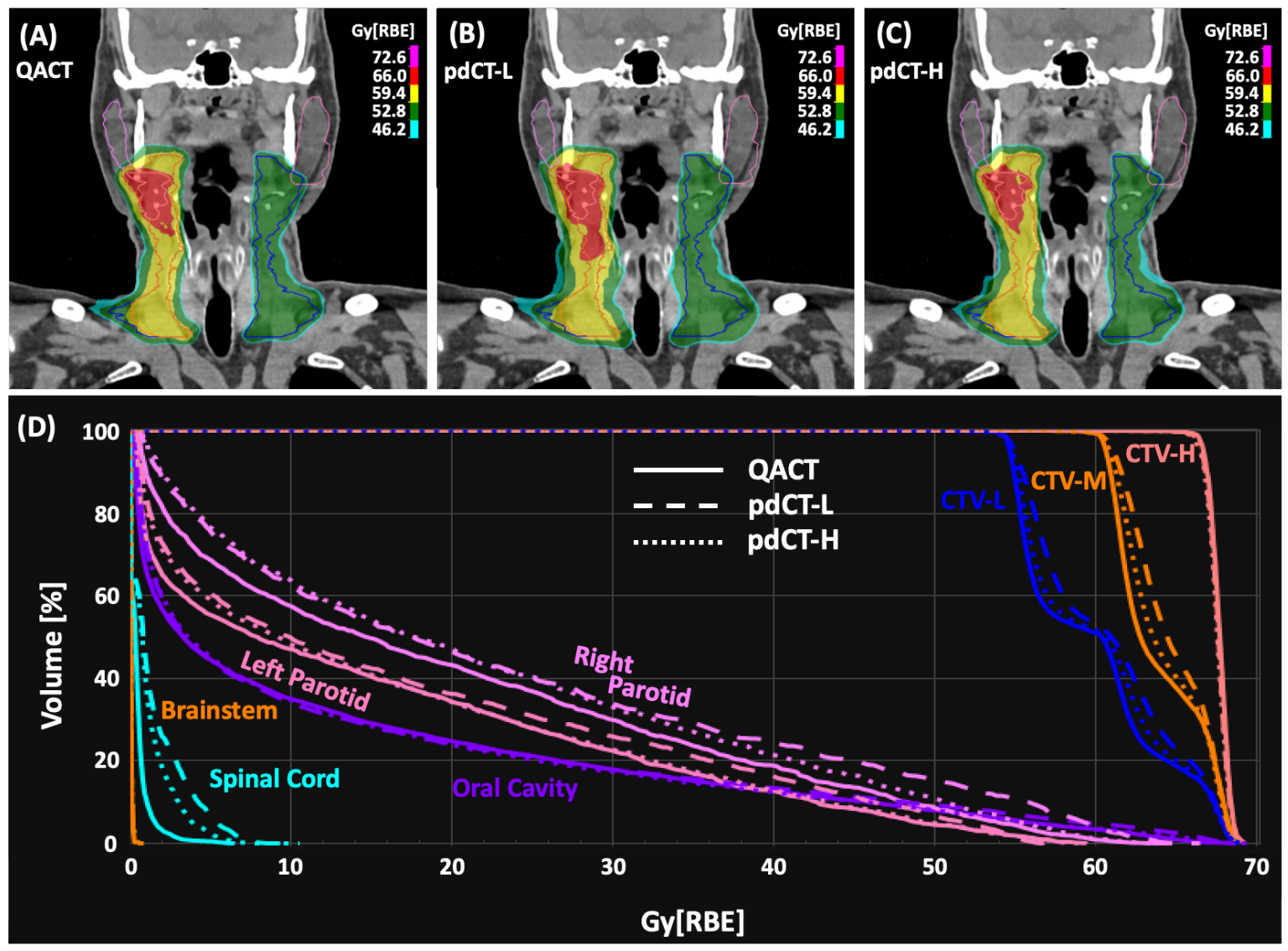

**Figure 5.** Dosimetric comparison of the offline replan and the online UGDT plans for patient P01. Coronal dose colorwash on the QACT for (A) the clinical offline replan, (B) the online adaptive plan initialized from pdCT-L and (C) the online adaptive plan initialized from pdCT-H; all three doses are computed on the same QACT. Isodose levels are given in the inset legend in Gy, and target and OAR contours are overlaid. (D) DVHs of the three plans for CTV-H, CTV-M, CTV-L, brainstem, spinal cord, oral cavity and the left and right parotid glands: solid lines, offline replan on the QACT; dashed lines, pdCT-L-initialized plan; dotted lines, pdCT-H-initialized plan. The target curves of the three plans are superimposed; the pdCT-L-initialized plan shifts the parotid curves toward higher dose, and both online plans deliver slightly more dose to the spinal cord than the offline replan.

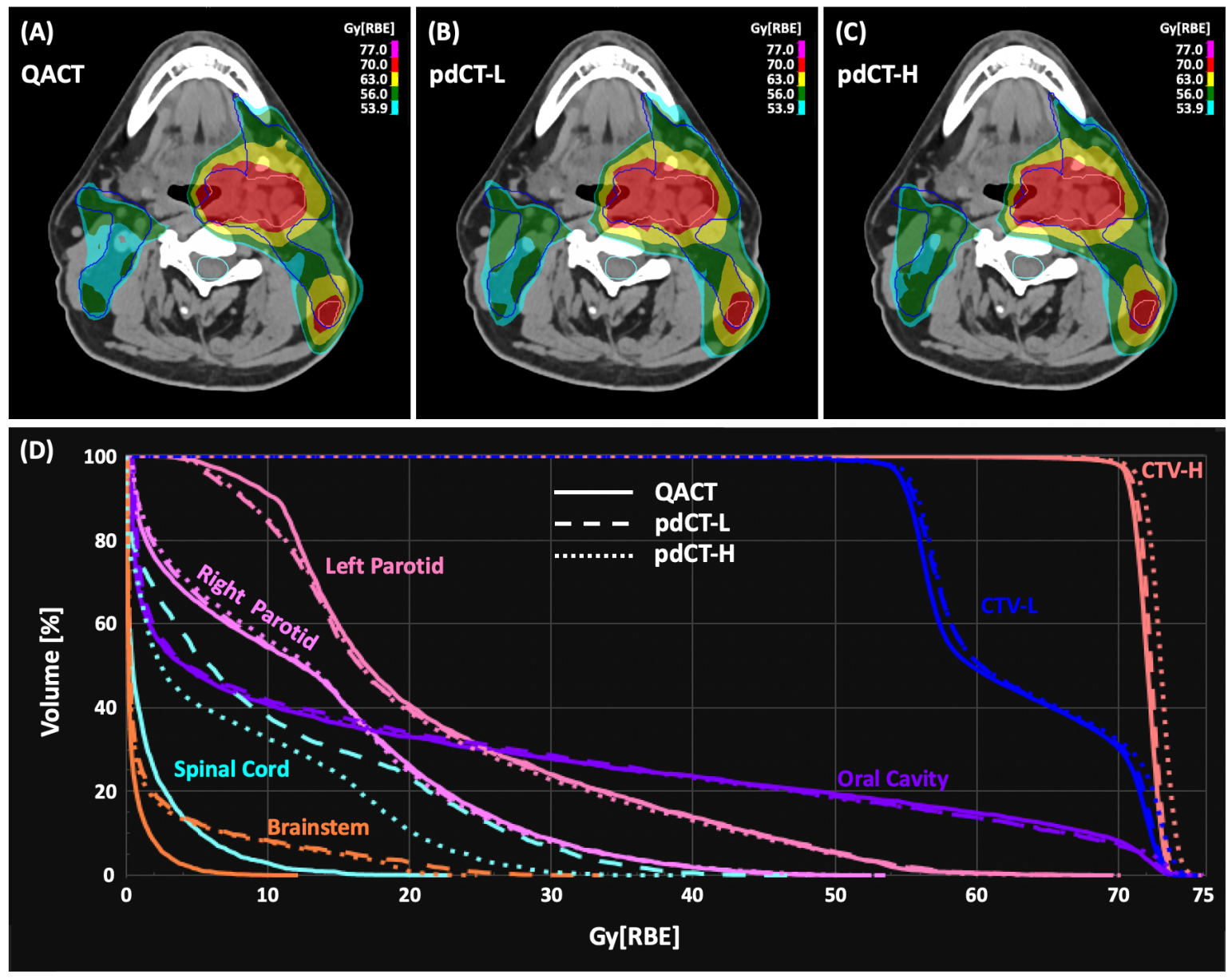

**Figure 6.** Dosimetric comparison of the offline replan and the online UGDT plans for patient P04. Axial dose colorwash on the QACT for (A) the clinical offline replan, (B) the online adaptive plan initialized from pdCT-L and (C) the online adaptive plan initialized from pdCT-H, with isodose levels in Gy in the inset legend. (D) DVHs for CTV-H, CTV-L, brainstem, spinal cord, oral cavity and the left and right parotid glands: solid, offline replan; dashed, pdCT-L-initialized plan; dotted, pdCT-H-initialized plan. The target, parotid and oral cavity curves of the three plans coincide; the spinal cord and brainstem curves of both online plans extend to higher dose than those of the offline replan, more so for the pdCT-L-initialized plan, while remaining below tolerance.

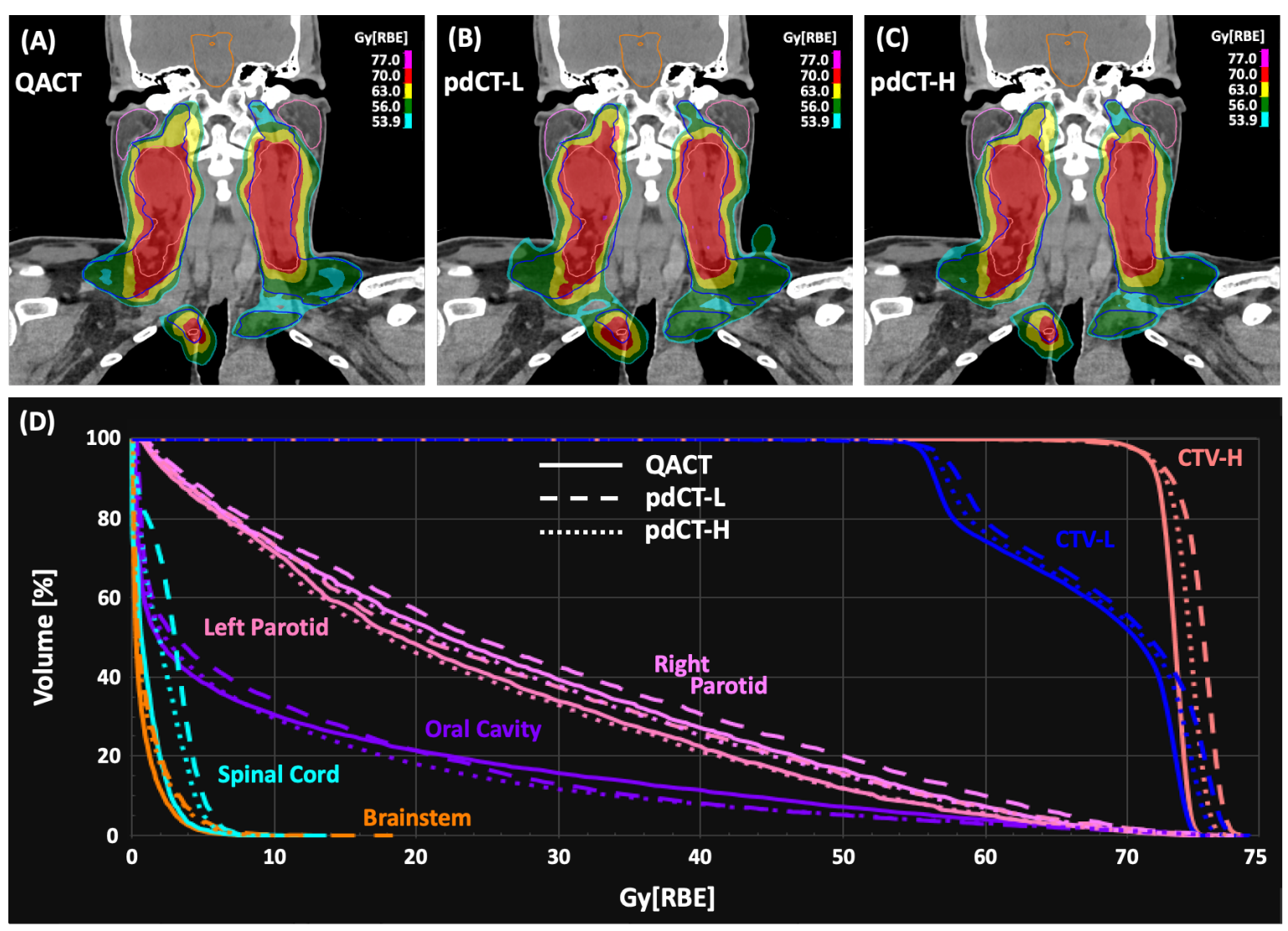

**Figure 7.** Dosimetric comparison of the offline replan and the online UGDT plans for patient P10. Coronal dose colorwash on the QACT for (A) the clinical offline replan, (B) the online adaptive plan initialized from pdCT-L and (C) the online adaptive plan initialized from pdCT-H, with isodose levels in Gy in the inset legend. (D) DVHs for CTV-H, CTV-L, brainstem, spinal cord, oral cavity and the left and right parotid glands: solid, offline replan; dashed, pdCT-L-initialized plan; dotted, pdCT-H-initialized plan. The pdCT-H-initialized plan reproduces the offline replan for the targets and for every OAR shown, whereas the pdCT-L-initialized plan shifts the parotid and brainstem curves toward higher dose and departs from the offline replan in the oral cavity.

## 4. Discussion

This work asked whether a UGDT framework that forecasts the treatment-day anatomy before the course begins can generate plans of clinical quality for HN online APT. The reference for clinical quality was the offline replan generated by a dosimetrist and approved by the treating physician on the QACT that triggered replanning, and the answer given by Tables 3 to 5 is affirmative within the limits of a ten-patient feasibility study. Plans built on the forecast anatomy scored within about 5% of that baseline, with nine of ten pdCT-H plans inside ±2% (Table 3). After reoptimization on the QACT, the online APT plans restored CTV-H D98% to 98.3 ± 0.3% against 98.5 ± 0.4% for the offline replan, kept the spinal cord and brainstem below tolerance in every patient, and scored within -1.1% of the baseline on average when initialized from pdCT-H, a difference that was not significant (Table 5). The plans initialized from the poorly matched pdCT-L reached the same target coverage but scored significantly lower than the baseline (-1.7%, $p = 0.01$), and were lower than the pdCT-H-initialized plans in seven of ten patients. The forecast anatomy therefore did what the framework requires of it: it supplied a clinically acceptable plan before the treatment day, and the better the forecast matched the anatomy of the day, the closer the adapted plan came to the clinical baseline.

The forward calculations of Table 4 show why the online reoptimization step is part of the framework rather than an option. Applied directly to the QACT, the pdCT plans lost 10.6 (pdCT-H) and 13.0 (pdCT-L) percentage points of CTV-H D98% on average, the spinal cord exceeded tolerance in two pdCT-L plans, and only four of twenty plans retained CTV-H D98% above 95%. OAR mean doses, by contrast, changed little between the predicted and the acquired anatomy. A forecast that matches the treatment-day image to a composite similarity of 0.95 is therefore close enough to plan on, but not close enough to deliver from, in a modality whose target coverage depends on the range of every pencil beam. The framework places the anatomical forecast, the contour-derived margin and the plan generation offline, and leaves for the treatment day a reoptimization that starts from a plan already tuned to an anatomy resembling the one presented. Table 5 shows that this reoptimization recovers target coverage completely, to a minimum CTV-H D98% of 98.0% across all twenty online plans.

The image-similarity ranking is a proxy for dosimetric usefulness, and the data allow it to be examined at three stages. At selection, the composite separated pdCT-H from pdCT-L by 0.019 to 0.029 in every patient, with all three terms lower for pdCT-L (Table 1). At forward calculation, pdCT-H retained more CTV-H coverage than pdCT-L on average, but the ordering reversed in four patients with differences of up to 22 percentage points (Table 4), so image similarity over the body does not order the coverage loss of an individual plan reliably. After reoptimization, the pdCT-H-initialized plans scored closer to the baseline in seven of ten patients, and only the pdCT-L set differed significantly from the baseline (Table 5). The ranking thus matters most where the framework uses it, at the initialization of the online plan. The residual difference between the online plans and the offline replan lies in the OARs rather than in the targets. Averaged over the cohort, the online plans placed 6.7 to 6.8 Gy more near-maximum dose in the spinal cord, 3.3 to 4.3 Gy more in the brainstem, and 1.7 to 3.0 Gy more mean dose in the oral cavity and parotid glands than the offline replan, and the same pattern was present in the initial pdCT plans (Table 3), so it originates in the plans generated on the forecast anatomy and is carried through reoptimization. The deficit is concentrated in a few patients: P03 accounts for the two largest score reductions, and P04 for the highest spinal cord doses, in both the initial and the online plans. Figures 5 to 7 locate the difference in the structures lying on steep dose gradients close to the targets, the spinal cord and brainstem in P04 and the parotid glands in P01 and P10, while the target DVHs of the three plans are superimposed. The online plans reused the objective template of the offline replan without further tuning and were optimized with the patient-specific margins of Table 2 as their setup uncertainty; which of these accounts for the OAR offset cannot be separated with the present data, and the offset kept eight of ten online plans within ±2% of the baseline score for either starting point.

The CTV robust margin is the uncertainty output of the framework. The margins of Table 2 ranged from 1.1 to 3.0 mm with a mean of 1.8 mm, were largest inferiorly and anteriorly and smallest superiorly, and differed between patients by nearly a factor of two, from 1.3 mm in P05 to 2.4 mm in P10. Figure 4 shows that this ordering follows the spread of the propagated contours: the patients with the widest ensembles, P01, P04, P06 and P08, received the largest margins, and those with the narrowest, P03 and P05, the smallest. The margin is therefore a property of how much anatomical change the library predicts for this patient, resolved by direction, and it is available before the first fraction from the same ensemble that produces the pdCTs. Whether a margin of this size is sufficient cannot be decided from a comparison against a replan that used its own clinical margins; the present data show only that plans optimized with these margins met target coverage on the treatment-day anatomy after reoptimization in all twenty cases.

Several limits qualify these findings. The cohort is ten patients from a single institution, each contributing one replanning QACT, so the results characterize feasibility and carry no claim about generalization or about the fractions between planning and replanning. The replanning QACT was chosen as the reference because it is the image on which the clinical baseline plan exists; it samples the course at the moment when anatomical change had become clinically consequential, and the framework has not been evaluated on days when it had not. The dose-volume endpoints on the QACT depend on the treatment-day contours drawn by the treating physician, and contour variability was not assessed. The plan quality score compresses many endpoints into one number, and a different scorecard could weigh the OAR offset differently. Within these limits, the present work establishes the precondition for DT-based online adaptation in proton therapy: the plans a UGDT framework generates for HN online APT reach the quality of physician-approved offline replans, and the framework opens a path for continued investigation of online APT toward personalized radiotherapy.

## 5. Conclusions

A UGDT framework for HN online APT was developed and evaluated in the dose domain. The framework forecasts the treatment-day anatomy before the course begins by transporting longitudinal deformations observed in previously treated patients onto the new patient's TPCT with a foundation model-based two-step multi-atlas DIR, derives a patient-specific, direction-resolved CTV robust margin from the contour uncertainty of the same ensemble, and selects the forecast closest to the treatment-day QACT by image similarity alone. In ten patients, plans generated on the forecast anatomy scored within about 5% of the physician-approved offline replan, and after online reoptimization on the QACT they matched the offline replan in target coverage and in spinal cord and brainstem tolerance, with a plan quality score within -1.1% of the baseline when initialized from the best-matching forecast. The framework therefore generates clinical-quality plans for a disease site with multiple prescription levels, many OARs and progressive anatomical change, and provides the anticipatory component that current reactive online adaptive workflows lack.

## Conflict of interest

The authors have no conflict of interest to disclose.

## Ethical statement

Institutional review board approval was obtained, and informed consent was not required for this Health Insurance Portability and Accountability Act compliant retrospective analysis. The research was conducted in accordance with the principles of the Declaration of Helsinki and with local statutory requirements.

## Data availability statement

The data cannot be made publicly available because they contain sensitive personal information. The data that support the findings of this study are available from the authors upon reasonable request.